\documentclass[10pt,letterpaper]{article}
\usepackage{opex3}
\usepackage{cite}

\begin{document}

\title{Characterization and Compensation of the Residual Chirp in a Mach-Zehnder-Type Electro-Optical Intensity Modulator}

\author{C. E. Rogers III, J. L. Carini, J. A. Pechkis, and P. L. Gould}

\address{Department of Physics \\University of Connecticut \\Storrs, CT 06269, USA}

\email{phillip.gould@uconn.edu} 


\begin{abstract}
     We utilize various techniques to characterize the residual phase modulation of a waveguide-based Mach-Zehnder electro-optical intensity modulator. A heterodyne technique is used to directly measure the phase change due to a given change in intensity, thereby determining the chirp parameter of the device. This chirp parameter is also measured by examining the ratio of sidebands for sinusoidal amplitude modulation. Finally, the frequency chirp caused by an intensity pulse on the nanosecond time scale is measured via the heterodyne signal. We show that this chirp can be largely compensated with a separate phase modulator. The various measurements of the chirp parameter are in reasonable agreement.
\end{abstract}



\section{Introduction} 

     Intensity modulators are important components of high-speed fiber-optic communication systems. They have also proven useful as fast switches in a variety of laser-based experiments in atomic and molecular physics. Of particular interest are the waveguide-based Mach-Zehnder-type electro-optic modulators. They have high speed, good extinction, and generally require low drive voltages. However, when used to modulate the intensity, there can be an accompanying residual phase modulation, or equivalently, a residual frequency chirp. The extent of this phase modulation for a given intensity modulation is quantified by the chirp parameter. For many applications, this residual chirp is undesirable. For example, in dense wavelength-division-multiplexing (DWDM) systems, chirp can lead to crosstalk between adjacent channels. In other situations, the chirp can be beneficial. For example, pulses with chirp have been used for adiabatic excitation/de-excitation in atom optics \cite{Miao2007,Bakos2007} and ultracold collision experiments \cite{Wright2007}. For these applications, excitation efficiencies can depend on the detailed temporal variations of frequency and intensity, a situation where the techniques of coherent control may be fruitfully applied.  In all cases, it is important to at least characterize, and possibly control, this chirp. Here we examine a particular intensity modulator and measure its chirp parameter using a variety of techniques. These rather distinct methods give consistent results. We also show that it is possible to largely compensate this residual chirp with a separate phase modulator.

	A number of techniques have been utilized to characterize the performance of Mach-Zehnder electro-optic intensity modulators and related devices, such as electroabsorption modulators. Sending the modulated light through a dispersive fiber, the frequency response was recorded with a network analyzer in order to obtain the chirp parameter \cite{Devaux1993,Schiess1994}. Stretched ultrafast pulses were modulated and then heterodyned with a delayed probe to measure the complex temporal response \cite{Park2007}. Frequency resolved optical gating has been used to characterize the intensity and phase properties of a high-speed modulator \cite{Thomsen2004}. A Mach-Zehnder modulator, driven by a delayed subharmonic, was employed to measure the chirp parameter of a separate modulated source \cite{Kowalski1999}. A Mach-Zehnder interferometer was utilized as an optical frequency discriminator to examine the chirp of a modulated source \cite{Saunders1994,Laverdière2003}. Specific modulation sidebands were selected with a tunable filter and their phase shifts compared to yield the chirp parameter \cite{Yan2003}. Analyzing the ratios of modulation sidebands, as we discuss in Sect. 4, has been used in a number of variations to determine the chirp parameter \cite{Auracher1980, Kawanishi2001, Oikawa2003, Courjal2004, Yan2005, Nagatsuka2007, Bakos2009}. Finally, various homodyne and heterodyne techniques have been employed to map out the amplitude and phase transfer functions of optical modulators \cite{Romstad2002, Krause2004, Dorrer2005}. Our technique is novel in that we use an optical heterodyne set-up to directly measure the phase shift as a function of modulation voltage under essentially static conditions. This yields directly the intrinsic chirp parameter. We also use the sideband ratio technique for comparison purposes. Finally, we measure via heterodyne the frequency chirp induced by a short intensity pulse and compare to the expected shape.
	
	The paper is organized as follows. In Sect. 2, we examine the operating principle of a Mach-Zehnder intensity modulator and the origin of residual phase modulation. In Sect. 3, we describe our heterodyne measurements of phase shift as a function of applied voltage. Our use of the sideband ratio technique is presented in Sect. 4. The short-pulse chirp measurements are discussed in Sect. 5. Sect. 6 comprises concluding remarks.

\section{Residual Phase Modulation and the Chirp Parameter} 

     The operating principle of a waveguide-based Mach-Zehnder-type electro-optic intensity modulator is shown in Fig. 1. Light from an optical fiber is coupled into a waveguide and then equally split into two paths which form the arms of a Mach-Zehnder interferometer. Light from these two arms is then recombined and coupled out to another fiber. The waveguides are made from lithium niobate, an electro-optic material, so that when a voltage is applied, a phase change is induced. To modulate the intensity, a controllable phase difference between the two arms is required. In the ideal intensity modulator, an equal but opposite phase would be induced in each arm, so that the phase of the output light is not modified. Such ``X-cut" devices are available, but generally have higher insertion loss and require somewhat higher drive voltages because of the larger electrode-waveguide spacing. A ``Z-cut" device, such as we use, has the electrode closer to one of the waveguides, causing an asymmetry between the two arms and therefore a phase modulation of the output when the intensity is modulated.

  \begin{figure}[htbp]
  \centering
  \includegraphics[width=9cm]{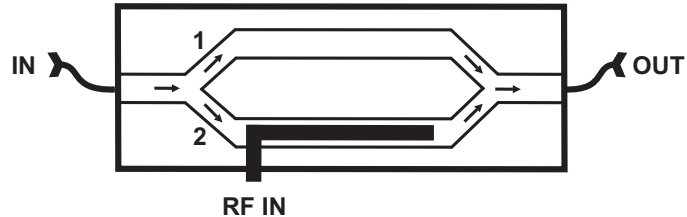}
  \caption{Schematic of Mach-Zehnder-type intensity modulator. The incident beam is equally split into two arms, 1 and 2, and recombined to provide the output. The phase difference between the two arms, controlled by the rf electrode, determines the output power via interference.}
  \end{figure}

	For our purposes, the important parameters of the Mach-Zehnder (MZ) intensity modulator (IM) are the voltage-to-phase conversion coefficients for the two arms, $\gamma_{1}$ and $\gamma_{2}$, which are assumed to be constant with respect to the applied modulation voltage V(t). Assuming that the input field of amplitude $E_{0}$ and frequency $\omega_{0}$ is equally split (without loss) between the two arms, the output field can be written as
	
\begin{equation}
\setcounter{equation}{1}
  E(t) = E_{1}(t) + E_{2}(t) = \frac{1} {2} E_{0}[e^{i(\omega_{0}t+\gamma_{1} \cdot V(t) + \varphi_{01})} + e^{i(\omega_{0}t+\gamma_{2} \cdot V(t) + \varphi_{02})}]
\end{equation}

Here, $\varphi_{01}$ and $\varphi_{02}$ are the static phases for each arm. With no modulation applied, a dc bias voltage controls the static phase difference $\Delta\varphi_{0}$ = $\varphi_{01}$ - $\varphi_{02}$ and thus determines the output level. In the presence of modulation, the output field can be expressed in terms of the time-dependent phase difference

\begin{equation}
\setcounter{equation}{2}
  \Delta\varphi(t) = \frac {1}{2} [(\gamma_{1} \cdot V(t)+ \varphi_{01}) - (\gamma_{2} \cdot V(t) + \varphi_{02})]
\end{equation}

and the time-dependent output phase

\begin{equation}
\setcounter{equation}{3}
  \varphi(t) = \frac {1}{2}[(\gamma_{1} \cdot V(t)+ \varphi_{01}) + (\gamma_{2} \cdot V(t) + \varphi_{02})]
\end{equation}

as

\begin{equation}
\setcounter{equation}{4}
  E(t) = E_{0}cos(\Delta\varphi(t))e^{i(\omega_{0}t + \varphi(t))}
\end{equation}

Since the device is an interferometer, the phase difference determines the ratio of output power to input power (assuming no loss)

\begin{equation}
\setcounter{equation}{5}
  \frac{P(t)}{P_{0}} = cos^{2}(\Delta\varphi(t))
\end{equation}

The voltage change required to go from minimum to maximum output power is given by $V_{\pi}$ = $\pi$/($\gamma_{1}$-$\gamma_{2}$). This corresponds to a change in $\Delta\varphi$ of $\pi$/2. The time-dependent frequency is the time derivative of the output phase

\begin{equation}
\setcounter{equation}{6}
  \omega(t) = \frac{d\varphi}{dt} = \frac{1}{2}(\gamma_{1}+\gamma_{2}) \cdot \frac{dV}{dt}
\end{equation}

As can be seen from Eqs. (5) and (6), if $\gamma_{2}$ = -$\gamma_{1}$, we have pure intensity modulation with no phase modulation, while if $\gamma_{2}$ = $\gamma_{1}$, we have pure phase modulation. Of course, in an actual device, the situation will be somewhere in between, as characterized by the intrinsic chirp parameter \cite{Kawanishi2001,Bakos2009}

\begin{equation}
\setcounter{equation}{7}
  \alpha_{0} = \frac{\gamma_{1}+\gamma_{2}}{\gamma_{1}-\gamma_{2}}
\end{equation}

This parameter is the ratio of the time derivative of the output phase $\varphi$(t) (responsible for phase modulation) to the time derivative of the phase difference    (responsible for intensity modulation). The case of $\alpha_{0}$ = 0 corresponds to pure intensity modulation, while $\alpha_{0}$ = $\infty$ corresponds to pure phase modulation. For modulation in only one arm of the MZ interferometer, $\alpha_{0}$ = 1. Note that this intrinsic chirp parameter $\alpha_{0}$ is not the same as the often-used intensity-dependent chirp parameter \cite{Koyama88}

\begin{equation}
\setcounter{equation}{8}
  \alpha = \frac{\frac{d\varphi}{dt}}{\frac{1}{E} \cdot \frac{dE}{dt}}
\end{equation}

In the specific case where the power is modulated with small amplitude about $P_{0}$/2, i.e., when $\Delta\varphi_{0}$ = -$\pi$/2, then $\alpha$ reduces to $\alpha_{0}$ \cite{Kawanishi2001}.

\section{Direct phase measurement} 

We use two main methods to characterize the residual phase modulation of the EO Space AZ-0K5-05-PFA-PFA-790 intensity modulator: optical heterodyne and spectral analysis. For both techniques, the light source is a 780 nm external-cavity diode laser (ECDL) \cite{Ricci1995} with a linewidth of $\sim$ 1 MHz. The heterodyne set-up used for the direct phase shift measurement is shown in Fig. 2. The idea is to combine the modulator output with a fixed frequency reference beam and measure the resulting beat signal on a 2 GHz photodiode (Thorlabs SV2-FC) connected to a 2 GHz digital oscilloscope (Agilent Infiniium 54852A DSO). The procedure consists of stepping the modulation voltage, which varies the output intensity, and measuring the voltage-dependent phase shift of the heterodyne signal. After each step, the voltage is fixed, so the frequency of the output light is equal to $\omega_{0}$. However, the phase of the output light varies with the modulation voltage (Eq. 3), so the phase of the heterodyne signal will shift after each step. The pattern of voltage steps is controlled with a Tektronix AFG 3252 240 MHz arbitrary waveform generator (AWG). The resulting intensity pattern is shown in Fig. 3a. Note that intensity is not directly proportional to voltage, as indicated in Eq. 5.

  \begin{figure}[htbp]
  \centering
  \includegraphics[width=7cm]{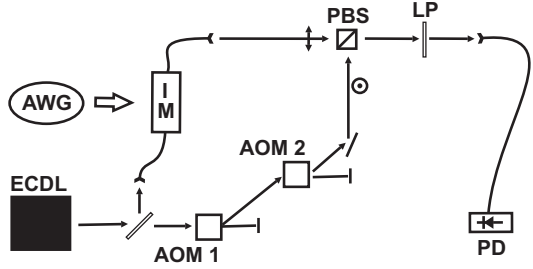}
  \caption{Set-up for measuring the output phase of the intensity modulator (IM) as the drive voltage is varied. The horizontally polarized output of the IM is combined on a polarizing beamsplitter cube (PBS) with a vertically polarized reference beam, which is derived from the input beam by frequency shifting a total of 160 MHz with two 80 MHz acousto-optical modulators (AOMs). The combined beams, whose relative intensities are adjusted with a linear polarizer (LP), produce a heterodyne signal on the photodiode (PD). The drive voltage is controlled with an arbitrary waveform generator (AWG).}
  \end{figure}

     The reference beam for the heterodyne is generated by frequency shifting the input light by a pair of acousto-optic modulators (AOMs) to give a beat frequency of 160 MHz. Deriving the reference beam from the modulator input has the advantage of making the heterodyne signal immune to common-mode frequency fluctuations. However, since we are measuring phase, we are sensitive to path length variations between the reference and signal beams. Therefore, the entire voltage waveform is completed in 2 $\mu$s, which is fast compared to time scale of vibrations and thermal drifts of optics in the heterodyne path. The time spent between voltage steps (approximately 130 ns) is many heterodyne periods, thus allowing an accurate determination of the phase. An example of a heterodyne signal, together with the corresponding sinusoidal fit, is shown in the inset of Fig. 3a. In order to further reduce the sensitivity to slow phase drifts, for the waveform shown in Fig. 3a, we return to the power level $P_{0}$/2  (V = $V_{\pi}$/2) after each step and measure phase shifts relative to the phase at this reference power (voltage) level. This local reference phase is determined by a single sinusoidal fit to the central 100 ns of the reference intervals before and after each interval of interest. In this fit, the amplitude, frequency, offset, and phase are free parameters. A similar fit is done in the interval of interest, but with the frequency now fixed at the value from the reference fit. The important parameter is the phase shift in the interval of interest. This phase shift as a function of voltage is shown in Fig. 3b. Since the abscissa of Fig. 3b is V/$V_{\pi}$ = $\pi$($\Delta\varphi$) (from Eq. 2), and the ordinate is the change in output phase $\varphi$ (from Eq. 3) divided by  $\pi$/2, the slope of this straight line gives directly the intrinsic chirp parameter $\alpha_{0}$. Averaging together the results from 45 repetitions, taken from three different voltage step patterns, we obtain $\alpha_{0}$ = 0.86(2), where the uncertainty is primarily statistical. Because the voltage waveform is completed in only 2 $\mu$s, and each phase measurement is sandwiched between two reference intervals, the uncertainty in each phase shift measurement is minimized. Variations in the reference phase are $<$0.01 rad for each phase shift measurement.

  \begin{figure}[htbp]
  \centering
  \includegraphics[width=10cm]{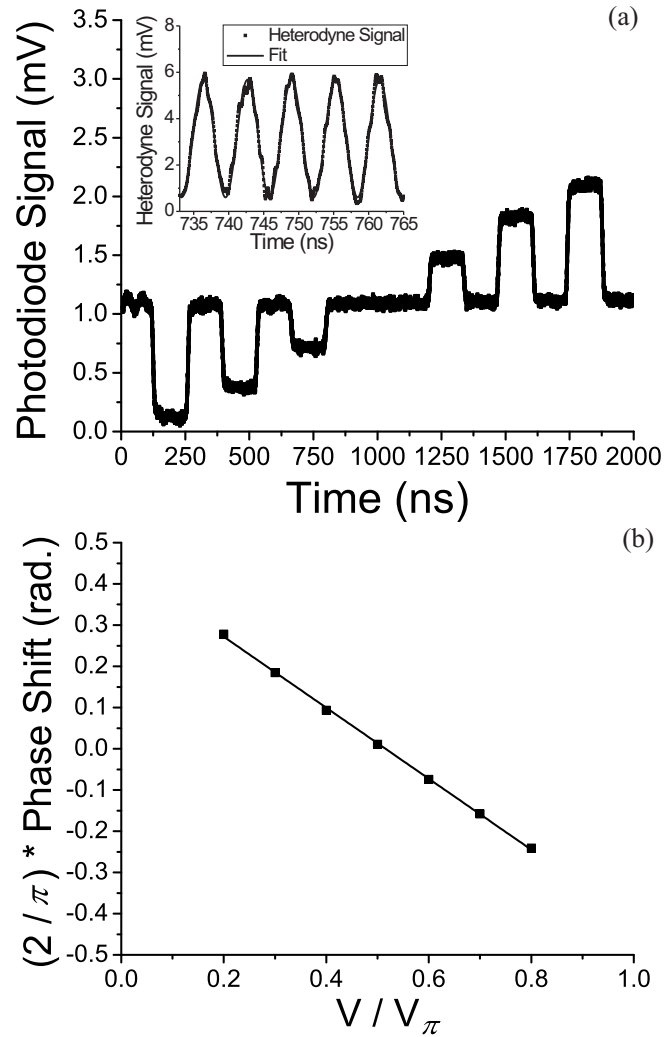}
  \caption{(a) Intensity waveform used to measure the phase shift as a function of IM drive voltage. This signal is generated by sending the output of the IM to the photodiode without the heterodyne reference beam. The inset shows a portion of the heterodyne signal and the sinusoidal fit used to determine the phase. (b) Phase shift of the IM output as a function of voltage, together with a linear fit. Uncertainties in the phase shifts are smaller than the points.}
  \end{figure}

     The value of $V_{\pi}$, which is needed for the determination of $\alpha_{0}$ discussed above, is obtained in a separate measurement. A slow, large amplitude sinusoidal modulation is applied to the modulator and the output power P is monitored. If the peak-to-peak voltage excursion is $V_{\pi}$, and the bias voltage is set to give $\Delta\varphi_{0}$ = $\pi$/2, then P swings between 0 and $P_{0}$. If the voltage excursion exceeds $V_{\pi}$, then P "wraps around" near its maxima and minima. For a peak-to-peak excursion of 2$V_{\pi}$, the output powers at the maximum and minimum voltages meet at $P_{0}$/2. If the bias voltage is slightly off from $\Delta\varphi_{0}$ = $\pi$/2, these output powers still meet at a common value. This method gives a rather precise measure of $V_{\pi}$ for two reasons: 1) reduced sensitivity to bias voltage; and 2) at the point where they match, the powers depend linearly on the voltage amplitude, so locating this point is easier than locating an extremum. Using this technique, we determine $V_{\pi}$ = 1.58(1) V at the slow modulation frequency of approximately 1 MHz. This is consistent with the specified value of 1.6 V at 1 GHz for our device.

\section{Determining the chirp parameter from modulation sidebands} 

The second method for measuring $\alpha_{0}$ involves sinusoidal modulation of the intensity and analysis of the resulting sidebands \cite{Auracher1980, Kawanishi2001, Oikawa2003, Courjal2004, Yan2005, Nagatsuka2007, Bakos2009}. Deviations of the sideband ratios from those expected for pure intensity modulation are indicative of residual phase modulation. Following the treatment of Bakos, et al. \cite{Bakos2009}, we apply sinusoidal modulation V = $V_{0}$sin($\omega$t) to Eq. 1 and obtain an output field

\begin{equation}
\setcounter{equation}{9}
  E(t) = \frac{1}{2}E_{0}e^{i\omega_{0} t}[e^{i(a_{1}sin(\omega t) + \varphi_{01})} + e^{i(a_{2} sin(\omega t) + \varphi_{02})}]
\end{equation}

where $a_{1,2}$ = $V_{0}$ $\cdot$ $\gamma_{1,2}$.  This can be Fourier decomposed into Bessel function sidebands

\begin{equation}
\setcounter{equation}{10}
  E(t) = \frac{1}{2}E_{0}e^{i\omega_{0} t}\displaystyle\sum_{n=-\infty}^{\infty}[J_{n}(a_{1})e^{i\varphi_{01}} + J_{n}(a_{2})e^{i\varphi_{02}}]e^{in\omega t}
\end{equation}

 For a fixed value of $\Delta\varphi_{0}$, set by the bias voltage, we can measure the intensity ratio of adjacent sidebands:

\begin{equation}
\setcounter{equation}{11}
  r_{n,n+1}=\frac{|J_{n}(a_{1})e^{i\varphi_{01}} + J_{n}(a_{2})e^{i\varphi_{02}}|^{2}}{|J_{n+1}(a_{1})e^{i\varphi_{01}} + J_{n+1}(a_{2})e^{i\varphi_{02}}|^{2}}
           =\frac{J_{n}^{2}(a_{1}) + J_{n}^{2}(a_{2}) + 2J_{n}(a_{1})J_{n}(a_{2})cos(\Delta\varphi_{0})}{J_{n+1}^{2}(a_{1}) + J_{n+1}^{2}(a_{2}) + 2J_{n+1}(a_{1})J_{n+1}(a_{2})cos(\Delta\varphi_{0})}
\end{equation}

We measure the intensities of the carrier (n = 0) and the first three sidebands (n = 1, 2, 3), thus obtaining three independent ratios. Since we only have two unknowns, $a_{1}$ and $a_{2}$, the system is over-determined. For simplicity, we take $\Delta\varphi_{0}$ = 0, the value we use in the experiment, and define the ratio $\beta$ = $a_{2}$/$a_{1}$ = $\gamma_{2}$/$\gamma_{1}$ so that Eq. 11 simplifies to

\begin{equation}
\setcounter{equation}{12}
  r_{n,n+1}=[\frac{J_{n}(a_{1})+J_{n}(\beta a_{1})}{J_{n+1}(a_{1})+J_{n+1}(\beta a_{1})}]^{2}
\end{equation}

For a given pair of sidebands, we define

\begin{equation}
\setcounter{equation}{13}
  \Delta_{n,n+1} = (J_{n}(a_{1}) + J_{n}(\beta a_{1}))^{2} - r_{n,n+1}(J_{n+1}(a_{1}) + J_{n+1}(\beta a_{1}))^{2}
\end{equation}

which should be equal to zero for the correct values of $a_{1}$ and $\beta$. To find these values, the three expressions for $\Delta_{n,n+1}$ (for n = 0, 1, 2) are plotted as functions of $a_{1}$ and $\beta$, and the common point where all three go to zero simultaneously is determined. Once $\beta$ is determined, we can use Eq. 7 to calculate the intrinsic chirp parameter:

\begin{equation}
\setcounter{equation}{14}
  \alpha_{0} = \frac{1+\beta}{1-\beta}
\end{equation}

	The measurement of sideband ratios is relatively straightforward. The modulator is biased at $\Delta\varphi_{0}$ = 0  (P = $P_{0}$) and a sinusoidal modulation of amplitude $V_{0}$ = 3 V and frequency $\omega$/(2$\pi$) = 240 MHz is applied. The spectrum is observed with a scanning Fabry-Perot interferometer (Coherent  33-6586-001) with 7.5 GHz free spectral range. A typical spectrum, showing only the carrier and first three positive sidebands, is shown in Fig. 4. Applying the above procedure to the measured sideband ratios yields $\alpha_{0}$ = 0.81(1), where the uncertainty is due mainly to the 5$\%$ uncertainty in measuring the height of each sideband. The sidebands are well resolved so crosstalk between them is negligible.
		
  \begin{figure}[htbp]
  \centering
  \includegraphics[width=7cm]{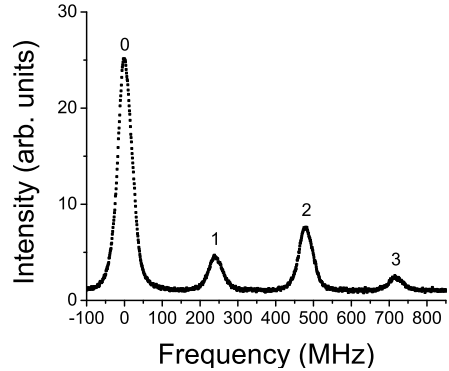}
  \caption{Optical spectrum for 240 MHz sinusoidal modulation of the IM. Only the carrier (0) and positive sidebands (1, 2, 3) are shown.}
  \end{figure}

\section{Chirp caused by an intensity pulse} 

	Since we are ultimately interested in using the intensity modulator to produce a specified pulse on the nanosecond time scale, we need to know the frequency chirp under these conditions. To measure the chirp, we use a variation of the heterodyne set-up described in Sect. 3. This is shown in Fig. 5. Since we are measuring on faster time scales, we need a higher beat frequency (e.g., 2 GHz), so we use a separate external-cavity diode laser for the reference beam. Also, for the chirp compensation discussed below, we add an electro-optic phase modulator (EO Space PM-0K1-00-PFA-PFA-790-S) prior to the intensity modulator.
	  	
  \begin{figure}[htbp]
  \centering
  \includegraphics[width=7cm]{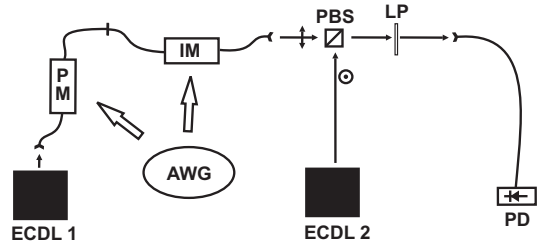}
  \caption{Set-up for measuring and compensating the frequency chirp in a pulse produced  by the IM. This is similar to Fig. 2, but with the incorporation of the phase modulator (PM) and the use of a separate laser for the heterodyne.}
  \end{figure}

	The voltage pulse driving the intensity modulator is generated by the AWG. To facilitate the diagnostics, we use a negative-going intensity pulse. This yields a strong heterodyne signal everywhere except at the very center of the pulse, allowing an accurate determination of the unshifted beat frequency. With a Gaussian voltage pulse programmed into the AWG, the resulting voltage output, together with the corresponding 2.55 ns FWHM Gaussian fit, are shown in Fig. 6a. Aside from some ringing on the trailing edge of the pulse due to the finite speed of the AWG, the fit is very good. Applying this voltage pulse to the intensity modulator yields the output intensity shown in Fig. 6b. Because the intensity modulator is an interferometer, and the phase difference in the two arms is proportional to the applied voltage, the output intensity is not directly proportional to voltage, as indicated in Eq. 5. Using the Gaussian derived from Fig. 6a, we fit this intensity pulse to Eq. 5. Because of timing delays between the electronic and optical signals, the centering of the Gaussian is allowed to be free parameter. The amplitude (fractional dip) is also a free parameter, but its value of 85.7\% is consistent with the value of 86.6\% predicted from Eq. 5, knowing the peak voltage from Fig. 6a and value of $V_{\pi}$. Once again, the overall fit is quite good. The electronic ringing seen in the voltage pulse is seen to carry through to the intensity pulse. Although this intensity pulse is not a Gaussian, such a pulse, or indeed any arbitrary pulse shape, can easily be realized by appropriately programming the AWG. Pulse widths are limited by the finite AWG bandwidth of 240 MHz (2 Gsamples/s).
	
  \begin{figure}[htbp]
  \centering
  \includegraphics[width=14cm]{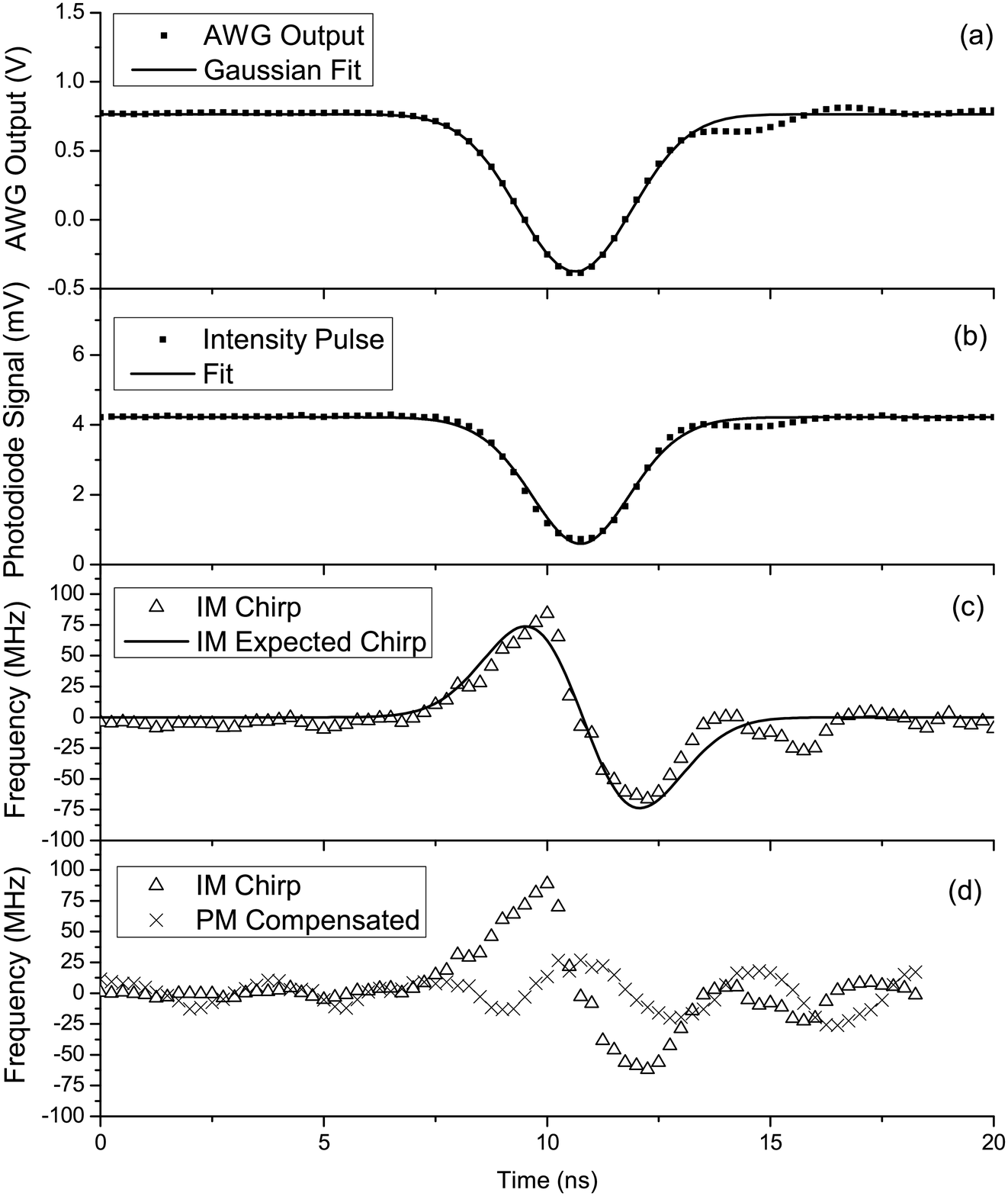}
  \caption{(a) Inverted Gaussian voltage pulse with fit. (b) Intensity pulse produced by voltage pulse in (a). The fit discussed in the text is also shown. (c) Measured frequency chirp (triangles) together with fit to derivative of the Gaussian (solid curve) assuming a chirp parameter of 0.86. (d) Frequency chirp due to IM (triangles) and with compensation by the PM (crosses). The chirp data is averaged over 150 repetitions.}
  \end{figure}
	
	The frequency chirp produced by the application of the Gaussian pulse to the intensity modulator is shown in Fig. 6c. The local frequency of the heterodyne signal is determined by measuring the period as the time interval between successive maxima and between successive minima. The overall heterodyne frequency of approximately 1.9 GHz, determined to within $\pm$2 MHz from the pre-pulse heterodyne signal, is subtracted from the measured frequencies. As mentioned above, this large offset, in conjunction with interpolation of the data, allows us to measure how the frequency is changing on a time scale significantly faster than the width of the pulse. Also shown in Fig. 6c (solid curve) is the chirp expected for the applied voltage pulse V(t) of Fig. 6a. Since the residual phase change is proportional to the voltage, and the frequency is the time derivative of the phase (Eq. 6), the frequency change will be proportional to the derivative of the Gaussian voltage pulse:

\begin{equation}
\setcounter{equation}{15}
  \omega (t) = \frac{\pi}{2} \cdot \alpha_{0} \cdot \frac{1}{V_{\pi}} \cdot \frac{dV}{dt}
\end{equation}

The solid curve in Fig. 6c is Eq. 15 with $V_{\pi}$ = 1.58 and $\alpha_{0}$ = 0.86, as determined in Sect. 3. Except for the ringing on the trailing edge, the agreement is quite good. The maximum chirp observed is -67 MHz/ns and the peak-to-peak frequency deviation is 151 MHz.

	In Fig. 6d, we demonstrate compensation of the residual IM frequency chirp using the phase modulator (PM). The PM is a similar device to the IM, but has only a single waveguide and is therefore not an interferometer. Its output phase is shifted in proportion to the input voltage, so we expect to be able to compensate residual phase modulation from the IM by applying the same voltage pulse shape to the PM. For the compensated curve in Fig. 6d, a Gaussian signal from a separate channel of the AWG is applied to the PM and its amplitude and width are adjusted to minimize the peak-to-peak frequency excursion. With a 2.55 ns FWHM pulse applied to the IM, the optimum PM pulse is slightly narrower, 2.37 ns FWHM. We believe that these small variations in pulse shape are due to the slightly different electrical responses of the IM and the PM. Using this technique, we are able to reduce the peak-to-peak frequency modulation by more than a factor of 3. Further reduction could likely be obtained by optimizing the shape of the signal applied to the PM through a genetic algorithm.

\section{Conclusion} 

We have investigated the residual frequency chirp from a Mach-Zehnder-type electro-optic intensity modulator. The most direct technique, using optical heterodyne to measure the output phase shift as a function of applied voltage, yields an intrinsic chirp parameter $\alpha_{0}$ = 0.86(2). A less direct measurement, based on sideband ratios for sinusoidal modulation, gives a slightly smaller value of $\alpha_{0}$ = 0.81(1). The first measurement is essentially at DC, while the second is at 240 MHz. Both of these values are marginally consistent with the value of 0.72(6) measured for a similar modulator at 450 MHz with the sideband technique \cite{Bakos2009}. Since $\alpha_{0}$ depends on the details of device fabrication, specifically on the electrode placement with respect to the two interferometer arms, it will not have a universal value. We have also examined the chirp resulting from the generation of a pulse with the intensity modulator. Heterodyne measurements show a chirp consistent with the measured parameters of the modulator. Using a separate phase modulator, we have demonstrated that the residual chirp can be partially compensated. Combining this type of intensity modulator with an arbitrary phase modulation system \cite{Rogers2007} will allow the production of arbitrary pulses with arbitrary chirps on the nanosecond time scale. This capability should prove useful for efficient excitation and coherent control in atomic and molecular systems.

\section{Acknowledgements} 

	This work was supported in part by the Chemical Sciences, Geosciences and Biosciences Division, Office of Basic Energy Sciences, U.S. Department of Energy. We thank EOSpace for technical advice regarding the intensity modulator.

\end{document}